\documentclass[aps,prd,onecolumn,showpacs,showkeys,amsmath,amssymb]{revtex4}
\usepackage{graphicx}
\usepackage{dcolumn}
\usepackage{amsmath}
\usepackage{bm}
\begin{document}
\title{Meson-baryon scattering lengths in HB$\chi$PT}
\author{Yan-Rui Liu}\email{yrliu@pku.edu.cn}
\affiliation{Department of Physics, Peking University, Beijing
100871, China}
\author{Shi-Lin Zhu}
\email{zhusl@th.phy.pku.edu.cn} \affiliation{Department of
Physics, Peking University, Beijing 100871, China}

\date{\today}

\begin{abstract}

We calculate the chiral corrections to the s-wave pseudoscalar
meson octet-baryon scattering lengths $a_{MB}$ to ${\cal O}(p^3)$
in the SU(3) heavy baryon chiral perturbation theory (HB$\chi$PT).
Hopefully the obtained analytical expressions will be helpful in
the chiral extrapolation of these scattering lengths in the future
lattice simulation.

\end{abstract}



\maketitle

\pagenumbering{arabic}

\section{Introduction}\label{sec1}

Heavy baryon chiral perturbation theory (HB$\chi$PT)
\cite{hbchpt,review} is widely used to study low energy processes
involving chiral fields and ground state baryons. Its Lagrangian is
expanded simultaneously with $p/\Lambda_\chi$ and $p/M_0$, where $p$
represents the meson momentum or its mass or the small residue
momentum of baryon in the non-relativistic limit, $\Lambda_\chi\sim
1\mathrm{GeV}$ is the scale of chiral symmetry breaking, and $M_0$
is the baryon mass in the chiral limit.

Pion-nucleon interactions are widely investigated in the
two-flavor HB$\chi$PT. For processes involving kaons or hyperons,
one has to employ the SU(3) HB$\chi$PT, where more unknown
low-energy constants (LEC) appear at the same chiral order than in
the SU(2) case . Determining these LECs needs more experimental
data which are unavailable at present. Broken SU(3) symmetry
results in large $m_K$ and $m_\eta$. In certain cases, the chiral
expansion converges slowly due to $m_K/\Lambda_\chi\sim
m_\eta/\Lambda_\chi\sim 1/2$ \cite{mei,lee,KN,axial}.

The scattering length is an important observable which contains
important information about low-energy meson-baryon strong
interactions. As an effective theory of QCD, HB$\chi$PT provides a
model-independent approach to calculate meson-baryon scattering
lengths. In this paper, we will calculate all s-wave meson-baryon
scattering lengths to the third chiral order in the SU(3)
HB$\chi$PT.

Experimental measurements of $\pi N$ scattering lengths are
relatively easier than those for other processes. Recently, the
results of precision X-ray experiments on pionic hydrogen
\cite{pihy} and deuterium \cite{pide} were published. These two
experiments together constrain the isoscalar and isovector $\pi N$
scattering lengths \cite{TpiN}.

Since the prediction of $\pi N$ scattering lengths with current
algebra, chiral corrections have been calculated to high orders in
the two-flavor HB$\chi$PT prior \cite{prsl,functions}. There were
many theoretical calculations of $\pi N$ scattering lengths
\cite{relsl}. As we will see below, $\pi N$ scattering lengths play
an important role in determining LECs in the SU(3) HB$\chi$PT.

Experimental data is very scarce for kaon-nucleon scattering
lengths. The recent DEAR measurements on kaonic hydrogen gave the
scattering length $a_{K^-p}$ \cite{DEAR}. However, the direct
determination of the two scattering lengths in the $\overline{K}N$
channel needs further experimental measurements. Unlike the
pion-nucleon case, there is the possibility that measurements on
kaonic deuterium are not enough to give the two $\overline{K}N$
scattering lengths \cite{meisana}.

There had been efforts to extract the scattering lengths from
kaon-nucleon scattering data using various models
\cite{TKN,TKNcom,VPI,cuni}. Kaon-nucleon scattering lengths were
calculated to order ${\cal O}(p^3)$ in the SU(3) HB$\chi$PT in Ref.
\cite{KN}. There were large cancellations at the second and the
third chiral order. Recently two lattice simulations also tried to
extract the kaon-nucleon scattering lengths \cite{latt1,latt2}.

$\eta N$ scattering is particularly interesting because the
attractive $\eta N$ interaction may result in $\eta$-mesic nuclei
which have a long history \cite{etamesic}. The formation of
$\eta$-mesic hypernuclei is also possible \cite{etahy}. An early
experiment gave a negative result on the search for $\eta$-mesic
bound states \cite{negative}. However, a recent experiment signals
the existence of such states \cite{mesicsig}. The experimental
situation is rather ambiguous.

On the theoretical side, the existence of $\eta$-$^3$He quasibound
state is also controversial, mostly because it is hard to
determine the $\eta$-nucleus scattering length. In the optical
models, the scattering length can be obtained with $\eta$N
scattering length $a_{\eta N}$. However, theoretical predictions
of $a_{\eta N}$ are different from various approaches (see
overview in \cite{etaNcompi}). It is worthwhile to calculate $\eta
N$ scattering length in the SU(3) HB$\chi$PT.

There are very few investigations of pion-hyperon, kaon-hyperon
and eta-baryon scattering lengths in literature. Experimentally
these observables may be studied through the strangeness program
at Japan Hadron Facility (JHF) and Lan-Zhou Cooling Storage Ring
(CSR). In this paper we will perform an extensive study of these
scattering lengths to the third chiral order in the framework of
SU(3) HB$\chi$PT.

We present the basic notations and definitions in Section
\ref{sec2}. We present the chiral corrections to the threshold
T-matrices of meson-baryon interactions in Section \ref{sec3}.
This section contains our main results. We discuss useful
relations between these threshold T-matrices in \ref{sec4}. Then
we discuss the determination of LECs in Section \ref{sec5}. The
final section is a summary.

\section{Lagrangian}\label{sec2}

For a system involving pions and one octet baryon, the chirally
invariant Lagrangian of HB$\chi$PT reads
\begin{equation}\label{lagr}
{\cal L}={\cal L}_{\phi\phi}+{\cal L}_{\phi B},
\end{equation}
where $\phi$ represents the pseudoscalar octet mesons, $B$
represents the octet baryons. The purely mesonic part ${\cal
L}_{\phi\phi}$ incorporates even chiral order terms while the
terms in ${\cal L}_{\phi B}$ start from ${\cal O}(p)$.
\begin{equation}
{\cal L}^{(2)}_{\phi\phi}=f^2 {\rm tr}(u_\mu
u^\mu+\frac{\chi_+}{4}),
\end{equation}
\begin{equation}
{\cal L}^{(1)}_{\phi B}={\rm tr}(\overline{B}(i\partial_0
B+[\Gamma_0,B]))-D {\rm tr}(\overline{B}\{\vec\sigma\cdot\vec
u,B\})-F {\rm tr}(\overline{B}[\vec\sigma\cdot \vec u,B]),
\end{equation}
\begin{eqnarray} {\cal L}_{\phi B}^{(2)} &=& b_D{\rm tr}(\overline B\{\chi_+,
B\} ) + b_F {\rm tr} (\overline B[\chi_+,B]) + b_0 {\rm
tr}(\overline B B){\rm tr} (\chi_+)  \nonumber \\
&&+ \Big(2d_D+{D^2 -3F^2 \over 2M_0}\Big) {\rm tr} (\overline B \{
u_0^2, B\} ) + \Big(2d_F-{DF\over M_0}\Big)  {\rm tr} (\overline B
[u_0^2,B]) \nonumber\\
&&+ \Big( 2d_0 +{F^2-D^2 \over 2M_0} \Big) {\rm tr}(\overline
BB){\rm tr} (u_0^2) \nonumber\\
&&+ \Big( 2d_1+{3F^2 -D^2 \over 3M_0} \Big) {\rm tr} (\overline B
u_0) {\rm tr}(u_0 B),
\end{eqnarray}
where
\begin{equation}
\Gamma_\mu = {i\over 2} [\xi^\dagger, \partial_\mu\xi],\qquad
u_\mu={i\over 2} \{\xi^\dagger, \partial_\mu \xi\},\qquad \xi =
\exp(i \phi/2f),
\end{equation}
\begin{equation}
\chi_+ = \xi^\dagger\chi\xi^\dagger+\xi\chi\xi,\qquad
\chi=\mathrm{diag}(m_\pi^2,\, m_\pi^2,\, 2m_K^2-m_\pi^2),
\end{equation}
\begin{eqnarray}
\phi=\sqrt2\left(
\begin{array}{ccc}
\frac{\pi^0}{\sqrt2}+\frac{\eta}{\sqrt6}&\pi^+&K^+\\
\pi^-&-\frac{\pi^0}{\sqrt2}+\frac{\eta}{\sqrt6}&K^0\\
K^-&\overline{K}^0&-\frac{2}{\sqrt6}\eta
\end{array}\right),\qquad
B=\left(
\begin{array}{ccc}
\frac{\Sigma^0}{\sqrt2}+\frac{\Lambda}{\sqrt6}&\Sigma^+&p\\
\Sigma^-&-\frac{\Sigma^0}{\sqrt2}+\frac{\Lambda}{\sqrt6}&n\\
\Xi^-&\Xi^0&-\frac{2}{\sqrt6}\Lambda
\end{array}\right).
\end{eqnarray}
$f$ is the meson decay constant in the chiral limit. $\Gamma_\mu$
is chiral connection which contains even number meson fields.
$u_\mu$ contains odd number meson fields. $D+F=g_A=1.26$ where
$g_A$ is the axial vector coupling constant. The first three terms
in ${\cal L}_{\phi B}^{(2)}$ are proportional to SU(3) symmetry
breaking. Terms involving $M_0$ are corrections produced by finite
baryon mass in the chiral limit. Others are proportional to LECs
$d_D$, $d_F$, $d_1$ and $d_0$.

We calculate threshold T-matrices for meson-baryon scattering to the
third order according to the power counting rule
\cite{hbchpt,review}. The leading and next leading order
contributions can be read from the tree level Lagrangians ${\cal
L}^{(1)}_{\phi B}$ and ${\cal L}^{(2)}_{\phi B}$ respectively.

At the third order, both loop diagrams and ${\cal O}(p^3)$ LECs
from ${\cal L}^{(3)}_{\phi B}$ contribute. In principle,
divergence from the loop integration will be absorbed by these
LECs. There are many unknown LECs at ${\cal O}(p^3)$ which may
reduce the predictive power of HB$\chi$PT. The contribution of
these ${\cal O}(p^3)$ LECs to threshold T-matrix was carefully
investigated using the resonance saturation approach in the SU(2)
HB$\chi$PT in Refs. \cite{prsl}. Luckily, the counter-term
contributions at ${\cal O}(p^3)$ are much smaller than the loop
corrections and the chiral corrections at this order mainly come
from the chiral loop \cite{prsl}. Therefore, these negligible
counter-term contributions were ignored in the calculation of
kaon-nucleon scattering lengths in Ref. \cite{KN}. We follow the
same approach in our crude numerical analysis because of the lack
of enough data to fix these small LECs.

\section{T-matrices for meson baryon scatterings}
\label{sec3}

The threshold T-matrix $T_{P B}$ is related to scattering length
$a_{PB}$ through $T_{P B}=4\pi(1+{m_P\over M_B}) a_{P B}$, where
$m_P$ and $m_B$ are the masses of the pseudoscalar meson and baryon,
respectively.

\begin{figure}
\begin{center}
\includegraphics{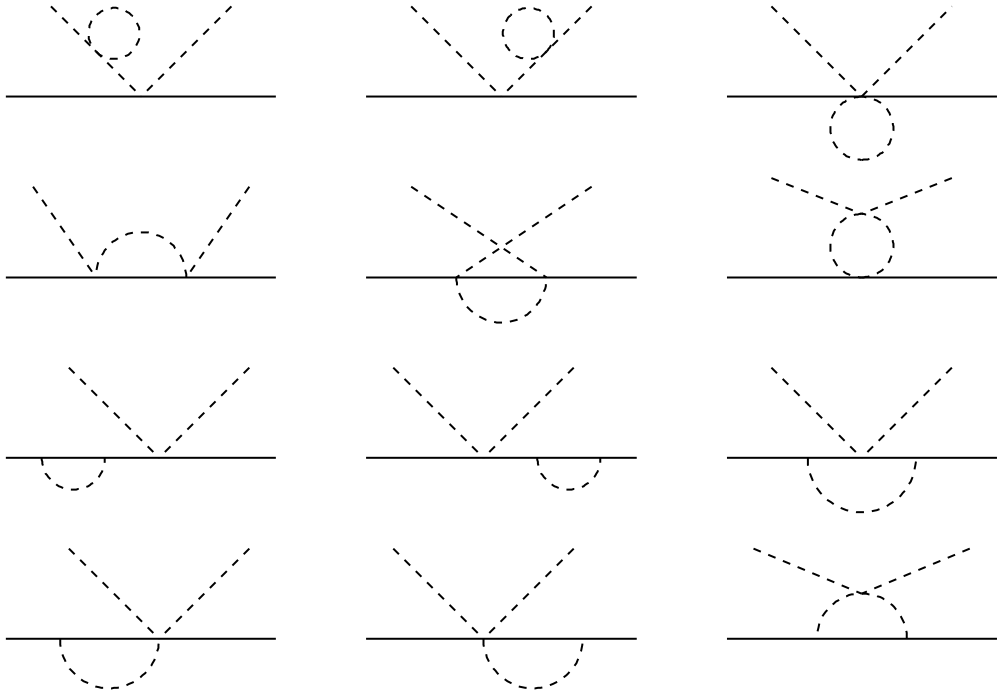}
\end{center}
\caption{Non-vanishing loop diagrams in the calculation of
meson-baryon scattering lengths to the third chiral order in
HB$\chi$PT. Dashed lines represent Goldstone bosons while solid
lines represent octet baryons. The fourth diagram generates
imaginary parts for kaon-baryon and eta-baryon scattering
lengths.}\label{MB}
\end{figure}

There exist many diagrams for a general elastic meson-baryon
scattering process. However, the calculation is simpler at
threshold due to $\vec{\sigma}\cdot\vec{q}=0$ where $\vec{q}$ is
the three momentum of the meson. The lowest order contribution
arises from the chiral connection term in ${\cal L}^{(1)}_{\phi
B}$. Only meson masses and decay constants appear in the
expressions. At the next leading order, the expressions are
proportional to the combinations of several LECs in ${\cal
L}^{(2)}_{\phi B}$. At the third chiral order, the loop diagrams
in Fig. \ref{MB} have non-vanishing contributions. The vertices
come from ${\cal L}^{(1)}_{\phi B}$ and ${\cal
L}^{(2)}_{\phi\phi}$

There are two types of diagrams. The first six contain vertices
from the chiral connection, and thus contributions from these
diagrams do not contain the axial coupling constants D and F.
Because the mass of an intermediate meson is always larger than or
equal to $m_\pi$, no diagrams generate imaginary parts for the
pion-baryon scattering lengths. In contrast, the fourth diagram
generates imaginary parts for kaon-baryon or eta-baryon threshold
T-matrices.

In the following subsections we list the expressions of threshold
T-matrices order by order for every channel. This is our main
result, which may be helpful to the chiral extrapolations of the
scattering lengths on the lattice.

\subsection{Kaon-nucleon scattering}

For kaon-nucleon scattering lengths, we reproduce the tree level
expressions of Ref. \cite{KN}. At the third chiral order, our
expressions are slightly different \footnote{In fact, we have
reached agreement on these loop corrections through communications
with Prof Kaiser.}.

At the leading order,
\begin{equation}
T_{KN}^{(1)}=-{m_K \over f_K^2}, \qquad T_{KN}^{(0)}=0, \qquad
T_{\overline KN}^ {(1)}={m_K  \over 2 f_K^2},  \qquad T_{\overline
KN}^{(0)}={3m_K\over 2f_K^2},
\end{equation}
where the superscripts represent total isospin and $f_K$ is the
renormalized kaon decay constant.

The second chiral order T-matrices are
\begin{eqnarray}
T_{KN}^{(1)} =\frac{m_K^2}{f_K^2}C_1, \qquad
T_{KN}^{(0)}=\frac{m_K^2}{f_K^2} C_0, \qquad T_{\overline
KN}^{(1)}=\frac{m^2_K}{2f_K^2}\bigg(C_1+C_0\bigg) , \qquad
T_{\overline KN}^{(0)}=\frac{m^2_K}{2f_K^2} \bigg(3C_1 -C_0\bigg),
\end{eqnarray}
where $C_{1,0}$ are defined in Ref. \cite{KN}
\begin{eqnarray}
C_1&=&2(d_0-2b_0)+2(d_D-2b_D)+d_1-{D^2+3F^2\over6M_0},\nonumber\\
C_0&=&2(d_0-2b_0)-2(d_F-2b_F)-d_1-\frac{D(D-3F)}{3M_0}.
\end{eqnarray}
We will express combinations of LECs in the other channels with
$C_0$ and $C_1$. This can reduce the number of parameters for a
given channel.

At the third order, we have
\begin{eqnarray}
T_{KN}^{(1)}&=&\frac{m_K^2}{16\pi^2 f_K^4}\bigg\{m_K
\bigg(-3+2\ln\frac{m_\pi}{\lambda} + \ln\frac{|m_K|}{\lambda}+3
\ln\frac{m_\eta}{\lambda} \bigg)  \nonumber \\
&& +2\sqrt{m_K^2-m_\pi^2} \ln\frac{m_K+\sqrt {m_K^2-m_\pi^2}}{m_\pi}
-3\sqrt{m_\eta^2-m_K^2}\arccos\frac{m_K}{m_\eta} \nonumber\\
&& - \frac{\pi}{6} (D-3F)\bigg[ 2(D+F) \frac{m_\pi^2}{m_\eta+m_\pi}
+(D+5F) m_\eta  \bigg] \bigg\} ,
\end{eqnarray}
\begin{eqnarray}
T_{KN}^{(0)}&=&\frac{3m_K^2}{16\pi^2 f_K^4}\bigg\{m_K \bigg(
\ln\frac{m_\pi}{\lambda}-\ln\frac{|m_K|}{\lambda}
\bigg) + \sqrt{m_K^2-m_\pi^2} \ln\frac{m_K+\sqrt{m_K^2-m_\pi^2}}{m_\pi}\nonumber\\
&& + \frac{\pi}{3} (D-3F) \bigg[(D+F) \frac{m_\pi^2}{m_\eta+m_\pi}
+\frac16(7D+3F) m_\eta \bigg] \bigg\} ,
\end{eqnarray}
\begin{eqnarray}
T_{\overline KN}^{(1)}&=&\frac{m_K^2}{32\pi^2 f_K^4} \bigg\{ m_K
\bigg(3-5\ln\frac{m_\pi}{\lambda} +2 \ln\frac{|m_K|}{
\lambda}-3 \ln\frac{m_\eta}{\lambda} \bigg)\nonumber  \\
&&+5\sqrt{m_K^2-m_\pi^2}\bigg( i\, \pi
-\ln\frac{m_K+\sqrt{m_K^2-m_\pi^2}}{m_\pi} \bigg) - 3\sqrt{m_\eta^2- m_K^2} \arccos\frac{-m_K}{m_\eta}  \nonumber \\
&& + \frac{\pi}{3} (D-3F) \bigg[ 2(D+F)\frac{m_\pi^2}{m_\eta+m_\pi}
+(3D-F)m_\eta \bigg] \bigg\} .
\end{eqnarray}
\begin{eqnarray}
T_{\overline KN}^{(0)}&=&\frac{3m_K^2}{32\pi^2 f_K^4} \bigg\{ m_K
\bigg(3- \ln\frac{m_\pi}{\lambda}-2\ln\frac{|m_K|}{
\lambda} -3 \ln\frac{m_\eta}{\lambda} \bigg) \nonumber\\
 &&+\sqrt{m_K^2-m_\pi^2} \bigg( i\,\pi
 -\ln\frac{m_K+\sqrt{m_K^2-m_\pi^2}}{m_\pi} \bigg) -3 \sqrt{m_\eta^2-m_K^2} \arccos\frac{-m_K}{m_\eta}
\nonumber \\
&& - \frac{\pi}{3} (D-3F) \bigg[2(D+F) \frac{m_\pi^2}{m_\eta+m_\pi}
+\frac13(5D+9F) m_\eta \bigg] \bigg\} .
\end{eqnarray}

In loop calculations, we have used dimensional regularization and
minimal subtraction scheme. $\lambda$ is the cutoff. In deriving the
expressions in square brackets, Gell-Mann$-$Okubo relation was used.
For the decay constants in the loop expressions, we have used $f_K$
in kaon processes in the above equations in order to make the
expressions more compact. One may also use $f_\pi$ in $\pi$ loops,
$f_K$ in kaon loops and $f_\eta$ in $\eta$ loops. The differences
are of high order.

\subsection{Pion-nucleon scattering}

With the isospin amplitude, the tree level expressions are
\begin{eqnarray}
T_{\pi N}^{(3/2)}=-\frac{m_\pi}{2f_\pi^2} \qquad T_{\pi
N}^{(1/2)}=\frac{m_\pi}{f_\pi^2},
\end{eqnarray}
and
\begin{eqnarray}
T_{\pi N}^{(3/2)}=\frac{m_\pi^2}{2f_\pi^2}\bigg(C_1+C_0+4C_\pi\bigg)
\qquad T_{\pi
N}^{(1/2)}=\frac{m_\pi^2}{2f_\pi^2}\bigg(C_1+C_0+4C_\pi\bigg),
\end{eqnarray}
where
\begin{equation}
C_\pi=(d_F-2b_F)-\frac{DF}{2M_0}\;.
\end{equation}
At the third order
\begin{eqnarray}
T_{\pi N}^{(3/2)}&=& \frac{m_\pi^2}{16\pi^2 f_\pi^4} \bigg\{ -m_\pi
\bigg( \frac32
-2\ln\frac{m_\pi}{\lambda}-\ln\frac{m_K}{\lambda}\bigg)\nonumber\\
&& -\sqrt{m_K^2-m_\pi^2}\bigg(\pi +
\arccos\frac{m_\pi}{m_K}\bigg)+\frac{\pi}{4}\bigg[3(D+F)^2m_\pi-\frac13(D-3F)^2m_\eta\bigg]\bigg\},
\end{eqnarray}
\begin{eqnarray}
T_{\pi N}^{(1/2)}&=& \frac{m_\pi^2}{16\pi^2 f_\pi^4} \bigg\{ 2m_\pi
\bigg( \frac32
-2\ln\frac{m_\pi}{\lambda}-\ln\frac{m_K}{\lambda}\bigg)\nonumber\\
&& -\sqrt{m_K^2-m_\pi^2}\bigg(\frac32\pi +
2\arcsin\frac{m_\pi}{m_K}\bigg)+\frac{\pi}{4}\bigg[3(D+F)^2m_\pi-\frac13(D-3F)^2m_\eta\bigg]\bigg\}.
\end{eqnarray}

The isospin even and isospin odd amplitudes are also widely used
in literature.
\begin{eqnarray}
T_{\pi N}^-&=&\frac{m_\pi}{2f_\pi^2}+\frac{m_\pi^2}{16\pi^2 f_\pi^4}
\bigg\{ m_\pi \bigg( \frac32
-2\ln\frac{m_\pi}{\lambda}-\ln\frac{m_K}{\lambda}\bigg)
-\sqrt{m_K^2-m_\pi^2}\arcsin\frac{m_\pi}{m_K}\bigg\},
\end{eqnarray}
\begin{eqnarray}
\qquad T_{\pi N}^+ &=&
\frac{m_\pi^2}{f_\pi^2}\bigg\{\frac12(C_1+C_0)+2C_\pi\bigg\}+\frac{3m_\pi^2}{64
\pi f_\pi^4} \bigg\{-2 \sqrt{m_K^2-m_\pi^2 } +(D+F)^2
m_\pi-\frac19(D-3F)^2 m_\eta \bigg\}.
\end{eqnarray}
The above expressions agree with those in Ref. \cite{KN}. The two
conventions are related through the following equations:
\begin{equation}
T_{\pi N}^{(3/2)}=T_{\pi N}^+ - T_{\pi N}^-, \qquad T_{\pi
N}^{(1/2)}=T_{\pi N}^+ + 2 T_{\pi N}^-.
\end{equation}
Sometimes a third convention with the isoscalar amplitude $T_{\pi
N}^0=T_{\pi N}^+$ and isovector amplitude $T_{\pi N}^1=-T_{\pi
N}^-$ are used in the literature.

Most of the threshold T-matrices for pion-baryon and kaon-baryon
scattering at the second order depend on the above three
combinations $C_1$, $C_0$ and $C_\pi$. If $C_1$, $C_0$ and $C_\pi$
can be fixed from $KN$ and $\pi N$ scattering lengths, many
predictions can be made.

\subsection{Pion-Sigma scattering}

T-matrices at the leading order:
\begin{eqnarray}
T_{\pi\Sigma}^{(2)}=-\frac{m_\pi}{f_\pi^2}, \qquad
T_{\pi\Sigma}^{(1)}=\frac{m_\pi}{f_\pi^2}, \qquad
T_{\pi\Sigma}^{(0)}=\frac{2m_\pi}{f_\pi^2}.
\end{eqnarray}
At the next leading order we have
\begin{eqnarray}
T_{\pi\Sigma}^{(2)}=\frac{m_\pi^2}{f_\pi^2}C_1, \qquad
T_{\pi\Sigma}^{(1)}=\frac{m_\pi^2}{f_\pi^2}\bigg(C_1-2C_d\bigg),
\qquad
T_{\pi\Sigma}^{(0)}=\frac{m_\pi^2}{f_\pi^2}\bigg(C_1+3C_d\bigg),
\end{eqnarray}
where
\begin{equation}
C_d=d_1-\frac{D^2-3F^2}{6M_0}.
\end{equation}
The extraction of $C_d$ needs additional inputs besides
pion-nucleon and kaon-nucleon scattering lengths. Due to the lack
of experimental data, one cannot determine it like the
determination of $C_{1,0,\pi}$. We will estimate its value when we
discuss numerical results.

At the third order, we have
\begin{eqnarray}
T_{\pi\Sigma}^{(2)}&=&\frac{m_\pi^2}{8\pi^2
f_\pi^4}\bigg\{-m_\pi\bigg(\frac32-2\ln\frac{m_\pi}{\lambda}-\ln\frac{m_K}{\lambda}\bigg)
-\sqrt{m_K^2-m_\pi^2}\arccos\frac{m_\pi}{m_K} +
\frac{\pi}{2}\bigg[3F^2 m_\pi-\frac13 D^2 m_\eta\bigg]\bigg\},
\end{eqnarray}
\begin{eqnarray}
T_{\pi\Sigma}^{(1)}&=&\frac{m_\pi^2}{8\pi^2
f_\pi^4}\bigg\{m_\pi\bigg(\frac32 - 2\ln\frac{m_\pi}{\lambda} -
\ln\frac{m_K}{\lambda} \bigg)
-\sqrt{m_K^2-m_\pi^2}\arccos\frac{-m_\pi}{m_K}
+\frac{\pi}{2}\bigg[(2D^2-3F^2) m_\pi-\frac13 D^2
m_\eta\bigg]\bigg\},
\end{eqnarray}
\begin{eqnarray}
T_{\pi\Sigma}^{(0)}&=&\frac{m_\pi^2}{8\pi^2
f_\pi^4}\bigg\{2m_\pi\bigg(\frac32 - 2\ln\frac{m_\pi}{\lambda} -
\ln\frac{m_K}{\lambda} \bigg)
-\sqrt{m_K^2-m_\pi^2}\bigg(\frac32\pi-2\arccos\frac{m_\pi}{m_K}\bigg)\nonumber\\
&&-\frac{\pi}{2}\bigg[3(D^2-4F^2) m_\pi+\frac13 D^2
m_\eta\bigg]\bigg\}.
\end{eqnarray}

\subsection{Pion-Cascade scattering}

Because both nucleon and cascade are isospin-1/2 baryons, one
expects similar expressions. At the leading order, the expressions
are the same as those for the $\pi N$ scattering,
\begin{eqnarray}
T_{\pi \Xi}^{(3/2)}=-\frac{m_\pi}{2f_\pi^2}, \qquad T_{\pi
\Xi}^{(1/2)}=\frac{m_\pi}{f_\pi^2}.
\end{eqnarray}

The next leading order contributions read
\begin{eqnarray}
T_{\pi \Xi}^{(3/2)}=\frac{m_\pi^2}{2f_\pi^2}\bigg(C_1+C_0 \bigg),
\qquad T_{\pi \Xi}^{(1/2)}=\frac{m_\pi^2}{2f_\pi^2}\bigg(C_1+C_0
\bigg).
\end{eqnarray}
The expressions involve only combinations of LECs in kaon-nucleon
scattering. This is the relic of isopin and U-spin symmetry (see
the relations in Section \ref{sec4}).

At the third order, we have
\begin{eqnarray}
T_{\pi \Xi}^{(3/2)}&=&\frac{m_\pi^2}{16\pi^2 f_\pi^4} \bigg\{ -m_\pi
\bigg( \frac32
-2\ln\frac{m_\pi}{\lambda}-\ln\frac{m_K}{\lambda}\bigg)\nonumber\\
&& -\sqrt{m_K^2-m_\pi^2}\bigg(\pi +
\arccos\frac{m_\pi}{m_K}\bigg)+\frac{\pi}{4}\bigg[3(D-F)^2
m_\pi-\frac13(D+3F)^2 m_\eta\bigg]\bigg\},
\end{eqnarray}
\begin{eqnarray}
T_{\pi \Xi}^{(1/2)}&=& \frac{m_\pi^2}{16\pi^2 f_\pi^4} \bigg\{2
m_\pi \bigg( \frac32
-2\ln\frac{m_\pi}{\lambda}-\ln\frac{m_K}{\lambda}\bigg)\nonumber\\
&& -\sqrt{m_K^2-m_\pi^2}\bigg(\frac12\pi +
2\arccos\frac{-m_\pi}{m_K}\bigg)+\frac{\pi}{4}\bigg[3(D-F)^2m_\pi-\frac13(D+3F)^2m_\eta\bigg]\bigg\}.
\end{eqnarray}
The expressions are similar to $\pi N$ T-matrices at this order.
The difference lies in the factor $F$. One can get these
expressions from those for $\pi N$ scattering through replacing
$F$ by $-F$. In fact, we can also get the second order expressions
from those of $T_{\pi N}$ through replacing $b_F$ by $-b_F$, $d_F$
by $-d_F$ and $F$ by $-F$.

\subsection{Pion-Lambda scattering}

The leading order contribution vanishes, which can be understood
through the crossing symmetry:
$T_{\pi^+\Lambda}=T_{\pi^-\Lambda}=[T_{\pi^+\Lambda}]_{m_\pi\rightarrow
-m_\pi}$. Thus $T_{\pi\Lambda}=-T_{\pi\Lambda}$ at the leading
order. This analysis is also available for the following
T-matrices, $T_{K\Lambda}$, $T_{\overline{K}\Lambda}$ and $T_{\eta
B}$ whose leading order contributions also vanish.

The tree-level contribution appears only at the next leading order
\begin{eqnarray}
T_{\pi\Lambda}&=&\frac{ m_\pi^2}{
3f_\pi^2}\bigg(C_1+2C_0+4C_\pi+C_d\bigg).
\end{eqnarray}

At the third order, for a $P\Lambda$ or $\eta B$ scattering, only
diagram 4, 5, 10, 11 and 12 in Fig. \ref{MB} have non-vanishing
contributions. The expression for $T_{\pi\Lambda}$ at this order
is
\begin{eqnarray}
T_{\pi\Lambda}=\frac{m_\pi^2}{16 \pi
f_\pi^4}\bigg\{-3\sqrt{m_K^2-m_\pi^2}+D^2\bigg(m_\pi-\frac13
m_\eta\bigg)\bigg\}.
\end{eqnarray}
The above expression was first derived in Ref. \cite{KN}.

\subsection{Kaon-Sigma scattering}

One expects the similarity between $T_{K\Sigma}$ and
$T_{\overline{K}\Sigma}$ because both $K$ and $\overline{K}$ are
isospin doublets. The leading order expressions are
\begin{eqnarray}
T_{K\Sigma}^{(3/2)}=-\frac{m_K}{2f_K^2}, \qquad
T_{K\Sigma}^{(1/2)}=\frac{m_K}{f_K^2}, \qquad
T_{\overline{K}\Sigma}^{(3/2)}=-\frac{m_K}{2f_K^2}, \qquad
T_{\overline{K}\Sigma}^{(1/2)}=\frac{m_K}{f_K^2}. \qquad
\end{eqnarray}

At the next leading order, we have
\begin{eqnarray}
&T_{K\Sigma}^{(3/2)}=\frac{m_K^2}{2f_K^2}\bigg(C_1+C_0+4C_\pi\bigg)
,\qquad T_{K\Sigma}^{(1/2)}=\frac{m_K^2}{2f_K^2}\bigg(C_1+C_0-
2C_\pi\bigg),& \nonumber\\
&T_{\overline{K}\Sigma}^{(3/2)}=\frac{m_K^2}{2f_K^2}
\bigg(C_1+C_0\bigg), \qquad
T_{\overline{K}\Sigma}^{(1/2)}=\frac{m_K^2}{2f_K^2}\bigg(C_1+C_0+6C_\pi\bigg)
.. &
\end{eqnarray}

At the third chiral order
\begin{eqnarray}
T_{K\Sigma}^{(3/2)}&=&\frac{m_K^2}{32\pi^2 f_K^4} \bigg\{- m_K
\bigg(3+\ln\frac{m_\pi}{\lambda} -4 \ln\frac{|m_K|}{
\lambda}-3 \ln\frac{m_\eta}{\lambda} \bigg)\nonumber  \\
&&+\sqrt{m_K^2-m_\pi^2}\bigg(2 i\, \pi
-\ln\frac{m_K+\sqrt{m_K^2-m_\pi^2}}{m_\pi} \bigg) -3\sqrt{m_\eta^2- m_K^2} \arccos\frac{m_K}{m_\eta}  \nonumber \\
&& + \frac{4}{3}\pi D \bigg[ 2F\frac{m_\pi^2}{m_\eta+m_\pi}
+(D+2F)m_\eta \bigg] \bigg\},
\end{eqnarray}
\begin{eqnarray}
 T_{K\Sigma}^{(1/2)}&=&\frac{m_K^2}{16\pi^2 f_K^4} \bigg\{ m_K
\bigg(3+\ln\frac{m_\pi}{\lambda} -4 \ln\frac{|m_K|}{
\lambda}-3 \ln\frac{m_\eta}{\lambda} \bigg)\nonumber  \\
&&+\sqrt{m_K^2-m_\pi^2}\bigg(\frac14 i\, \pi
+\ln\frac{m_K+\sqrt{m_K^2-m_\pi^2}}{m_\pi} \bigg) -3\sqrt{m_\eta^2- m_K^2}\bigg(\frac34 \pi- \arccos\frac{m_K}{m_\eta}\bigg)  \nonumber \\
&& + \frac{2}{3}\pi D \bigg[ -4F\frac{m_\pi^2}{m_\eta+m_\pi}
+(D-4F)m_\eta \bigg] \bigg\},
 \end{eqnarray}
\begin{eqnarray}
T_{\overline{K}\Sigma}^{(3/2)}&=&\frac{m_K^2}{32\pi^2 f_K^4}
\bigg\{- m_K \bigg(3+\ln\frac{m_\pi}{\lambda} -4 \ln\frac{|m_K|}{
\lambda}-3 \ln\frac{m_\eta}{\lambda} \bigg)\nonumber  \\
&&+\sqrt{m_K^2-m_\pi^2}\bigg(2 i\, \pi
-\ln\frac{m_K+\sqrt{m_K^2-m_\pi^2}}{m_\pi} \bigg) -3\sqrt{m_\eta^2- m_K^2} \arccos\frac{m_K}{m_\eta}  \nonumber \\
&& + \frac{4}{3}\pi D \bigg[ -2F\frac{m_\pi^2}{m_\eta+m_\pi}
+(D-2F)m_\eta \bigg] \bigg\},
\end{eqnarray}
\begin{eqnarray}
T_{\overline{K}\Sigma}^{(1/2)}&=&\frac{m_K^2}{16\pi^2 f_K^4} \bigg\{
m_K \bigg(3+\ln\frac{m_\pi}{\lambda} -4 \ln\frac{|m_K|}{
\lambda}-3 \ln\frac{m_\eta}{\lambda} \bigg)\nonumber  \\
&&+\sqrt{m_K^2-m_\pi^2}\bigg(\frac14 i\, \pi
+\ln\frac{m_K+\sqrt{m_K^2-m_\pi^2}}{m_\pi} \bigg) -3\sqrt{m_\eta^2- m_K^2}\bigg(\frac34 \pi- \arccos\frac{m_K}{m_\eta}\bigg)  \nonumber \\
&& + \frac{2}{3}\pi D \bigg[ 4F\frac{m_\pi^2}{m_\eta+m_\pi}
+(D+4F)m_\eta \bigg] \bigg\}.
\end{eqnarray}
$T_{\overline{K}\Sigma}^{(3/2)}$ differs from
$T_{K\Sigma}^{(3/2)}$ in the sign in front of $F$ at the third
order. The same property holds for
$T_{\overline{K}\Sigma}^{(1/2)}$ and $T_{K\Sigma}^{(1/2)}$. One
can also testify that differences between $T_{K\Sigma}$ and
$T_{\overline{K}\Sigma}$ at the second order are the signs in
front of $b_F$, $d_F$ and $F$.

\subsection{Kaon-Cascade scattering}

The leading order contributions are
\begin{eqnarray}
T_{K\Xi}^{(1)}=\frac{m_K}{2f_K^2}, \qquad
T_{K\Xi}^{(0)}=\frac{3m_K}{2f_K^2},\qquad
T_{\overline{K}\Xi}^{(1)}=-\frac{m_K}{f_K^2}, \qquad
T_{\overline{K}\Xi}^{(0)}=0.
\end{eqnarray}

At the second order, we have
\begin{eqnarray}
&T_{K\Xi}^{(1)}=\frac{m_K^2}{2f_K^2}\bigg(C_1+C_0+4C_\pi\bigg)
,\qquad T_{K\Xi}^{(0)}=\frac{m_K^2}{2f_K^2}\bigg(3C_1-C_0-
4C_\pi\bigg),& \nonumber\\
&T_{\overline{K}\Xi}^{(1)}=\frac{m_K^2}{f_K^2} C_1, \qquad
T_{\overline{K}\Xi}^{(0)}=\frac{m_K^2}{f_K^2}\bigg(C_0+4C_\pi\bigg).&
\end{eqnarray}

The third order expressions are
\begin{eqnarray}
T_{K\Xi}^{(1)}&=&\frac{m_K^2}{32\pi^2 f_K^4} \bigg\{ m_K
\bigg(3-5\ln\frac{m_\pi}{\lambda} +2 \ln\frac{|m_K|}{
\lambda}-3 \ln\frac{m_\eta}{\lambda} \bigg)\nonumber  \\
&&+5\sqrt{m_K^2-m_\pi^2}\bigg( i\, \pi
-\ln\frac{m_K+\sqrt{m_K^2-m_\pi^2}}{m_\pi} \bigg) -3\sqrt{m_\eta^2- m_K^2} \arccos\frac{-m_K}{m_\eta}  \nonumber \\
&& + \frac{\pi}{3} (D+3F) \bigg[ 2(D-F)\frac{m_\pi^2}{m_\eta+m_\pi}
+(3D+F)m_\eta \bigg] \bigg\},
\end{eqnarray}
\begin{eqnarray}
 T_{K\Xi}^{(0)}&=&\frac{3m_K^2}{32\pi^2 f_K^4} \bigg\{ m_K
\bigg(3- \ln\frac{m_\pi}{\lambda}-2\ln\frac{|m_K|}{
\lambda} -3 \ln\frac{m_\eta}{\lambda} \bigg) \nonumber\\
 &&+\sqrt{m_K^2-m_\pi^2} \bigg( i\,\pi
 -\ln\frac{m_K+\sqrt{m_K^2-m_\pi^2}}{m_\pi} \bigg)  -3 \sqrt{m_\eta^2-m_K^2} \arccos\frac{-m_K}{m_\eta}
\nonumber \\
&& - \frac{\pi}{3} (D+3F) \bigg[2(D-F) \frac{m_\pi^2}{m_\eta+m_\pi}
+\frac13(5D-9F) m_\eta \bigg] \bigg\},
\end{eqnarray}
\begin{eqnarray}
T_{\overline{K}\Xi}^{(1)}&=&\frac{m_K^2}{16\pi^2 f_K^4}\bigg\{m_K
\bigg(-3+2\ln\frac{m_\pi}{\lambda} + \ln\frac{|m_K|}{\lambda}+3
\ln\frac{m_\eta}{\lambda} \bigg)  \nonumber \\
&& +2\sqrt{m_K^2-m_\pi^2} \ln\frac{m_K+\sqrt {m_K^2-m_\pi^2}}{m_\pi}  -3\sqrt{m_\eta^2-m_K^2}\arccos\frac{m_K}{m_\eta} \nonumber\\
&& - \frac{\pi}{6} (D+3F)\bigg[ 2(D-F) \frac{m_\pi^2}{m_\eta+m_\pi}
+(D-5F) m_\eta  \bigg] \bigg\},
\end{eqnarray}
\begin{eqnarray}
T_{\overline{K}\Xi}^{(0)}&=&\frac{3m_K^2}{16\pi^2 f_K^4}\bigg\{m_K
\bigg( \ln\frac{m_\pi}{\lambda}-\ln\frac{|m_K|}{\lambda}
\bigg) + \sqrt{m_K^2-m_\pi^2} \ln\frac{m_K+\sqrt{m_K^2-m_\pi^2}}{m_\pi}\nonumber\\
&& + \frac{\pi}{3} (D+3F) \bigg[(D-F) \frac{m_\pi^2}{m_\eta+m_\pi}
+\frac16(7D-3F) m_\eta \bigg] \bigg\} .
\end{eqnarray}

\subsection{Kaon-Lambda scattering}

Non-vanishing contributions start from the next leading order.
\begin{eqnarray}
&T_{\overline{K}\Lambda}=T_{K\Lambda}=\frac{m_K^2}{6f_K^2}\bigg[5C_1+C_0+2C_\pi-4C_d\bigg].&
\end{eqnarray}
At the third chiral order loop corrections are
\begin{eqnarray}
T_{\overline{K}\Lambda}=T_{K\Lambda}=\frac{9m_K^2}{64\pi
f_K^4}\bigg\{i\sqrt{m_K^2-m_\pi^2}-\sqrt{m_\eta^2-m_K^2}+\frac{8}{27}D^2m_\eta\bigg\}.
\end{eqnarray}
Similarly, one may get $T_{\overline{K}\Lambda}$ from
$T_{K\Lambda}$ through the replacement $b_F\rightarrow -b_F$,
$d_F\rightarrow -d_F$ and $F\rightarrow -F$. However, $b_F$ and
$d_F$ disappear in the second order T-matrices and $F$ disappears
in the third order T-matrix. Hence
$T_{\overline{K}\Lambda}=T_{K\Lambda}$.

\subsection{Eta-baryon scattering}

The leading order contribution vanishes for every channel. With
the Gell-Mann$-$Okubo relation the next leading order
contributions can be written as:
\begin{eqnarray}
T_{\eta N}&=&\frac{1}{6f^2_\eta}\bigg[(5C_1+C_0-4C_\pi-4C_d)m_\eta^2-4(b_D-3b_F)(m_\eta^2-m_\pi^2)\bigg],\nonumber\\
T_{\eta\Sigma}&=&\frac{1}{3f_\eta^2}\bigg[(C_1+2C_0+4C_\pi+C_d)m_\eta^2+4b_D(m_\eta^2-m_\pi^2)\bigg],\nonumber\\
T_{\eta\Xi}&=&\frac{1}{6f^2_\eta}\bigg[( 5C_1+C_0
+8C_\pi-4C_d)m_\eta^2-4(b_D+3b_F)(m_\eta^2-m_\pi^2)\bigg],\nonumber\\
T_{\eta\Lambda}&=&\frac{1}{3f^2_\eta}\bigg[ 3(C_1+ C_d)m_\eta^2-4b_D
(m_\eta^2-m_\pi^2)\bigg].
\end{eqnarray}
These expressions rely on $C_{1,0,\pi,d}$ as well as $b_D$ and
$b_F$. $b_{D}$ and $b_F$ are determined with the octet baryon
masses.

At the third chiral order, we have
\begin{eqnarray}
T_{\eta N}&=&\frac{1}{32\pi f_\eta^4}\bigg\{9i
m_\eta^2\sqrt{m_\eta^2-m_K^2}-\frac16(D-3F)^2(4m_\eta^2-m_\pi^2)m_\eta\nonumber\\
&&-\frac32(D+F)^2m_\pi^3+\frac23(5D^2+9F^2-6DF)m_K^3\bigg\},
\end{eqnarray}
\begin{eqnarray}
T_{\eta\Sigma}&=&\frac{1}{16\pi f_\eta^4}\bigg\{3i
m_\eta^2\sqrt{m_\eta^2-m_K^2}-\frac13 D^2(4m_\eta^2-m_\pi^2)m_\eta\nonumber\\
&&-\frac13(D^2+6F^2)m_\pi^3+2(D^2+F^2)m_K^3\bigg\},
\end{eqnarray}
\begin{eqnarray}
T_{\eta\Xi}&=&\frac{1}{32\pi f_\eta^4}\bigg\{9i
m_\eta^2\sqrt{m_\eta^2-m_K^2}-\frac16(D+3F)^2(4m_\eta^2-m_\pi^2)m_\eta\nonumber\\
&&-\frac32(D-F)^2m_\pi^3+\frac23(5D^2+9F^2+6DF)m_K^3\bigg\},
\end{eqnarray}
\begin{eqnarray}
T_{\eta\Lambda}&=&\frac{1}{16\pi f_\eta^4}\bigg\{9i
m_\eta^2\sqrt{m_\eta^2-m_K^2}-\frac13 D^2(4m_\eta^2-m_\pi^2)m_\eta
-D^2m_\pi^3+\frac23(D^2+9F^2)m_K^3\bigg\}.
\end{eqnarray}
Again there is the similarity between $T_{\eta N}$ and
$T_{\eta\Xi}$.

\section{Threshold T-matrix Relations}\label{sec4}

In this section, we list the relations between kaon-baryon and
anti-kaon-baryon threshold T-matrices according to the crossing
symmetry. We also present the relations of these T-matrices in the
SU(3) symmetry limit. These relations are used to cross-check our
calculations.

Crossing symmetry relates $KB$ processes with $\overline{K}B$
processes. According to this symmetry, we have the following
relations:
\begin{equation}
T_{\overline KN}^{(1)} = {1\over 2} \Big[ T_{KN}^{(1)}+
T_{KN}^{(0)}\Big]_{m_K\to -m_K}, \qquad T_{\overline KN}^{(0)} =
{1\over 2} \Big[ 3T_{KN}^{(1)}- T_{KN}^{(0)}\Big]_{m_K\to -m_K},
\end{equation}
\begin{equation}
T_{\overline K\Xi}^{(1)} = {1\over 2} \Big[ T_{K\Xi}^{(1)}+
T_{K\Xi}^{(0)}\Big]_{m_K\to -m_K}, \qquad T_{\overline K\Xi}^{(0)} =
{1\over 2} \Big[ 3T_{K\Xi}^{(1)}- T_{K\Xi}^{(0)}\Big]_{m_K\to -m_K},
\end{equation}
\begin{equation}
T_{\overline K\Sigma}^{(3/2)} = {1\over 3} \Big[
T_{K\Sigma}^{(3/2)}+ 2T_{K\Sigma}^{(1/2)}\Big]_{m_K\to -m_K}, \qquad
T_{\overline K\Sigma}^{(1/2)} = {1\over 3} \Big[
4T_{K\Sigma}^{(3/2)}- T_{K\Sigma}^{(1/2)}\Big]_{m_K\to -m_K},
\end{equation}
and
\begin{equation}
T_{\overline K\Lambda}=\Big[T_{K\Lambda}\Big]_{m_K\to -m_K}.
\end{equation}

In the SU(3) flavor symmetry limit, we derive the following
relations using the isospin and U-spin symmetry:
\begin{eqnarray}
&T_{KN}^{(1)}=T_{\pi\Sigma}^{(2)}=T_{\overline{K}\Xi}^{(1)},&\nonumber\\
&T_{\pi N}^{(3/2)}=T_{K\Sigma}^{(3/2)}=\frac12[
T_{\overline{K}\Xi}^{(1)} + T_{\overline{K}\Xi}^{(0)}],&\nonumber\\
&T_{\pi \Xi}^{(3/2)}=T_{\overline{K}\Sigma}^{(3/2)}=\frac12[
T_{KN}^{(1)} + T_{KN}^{(0)}],&\nonumber\\
&T_{K\Xi}^{(1)}=\frac13[T_{\pi N}^{(3/2)}+ 2T_{\pi N}^{(1/2)}]
=\frac13[ T_{\overline{K}\Sigma}^{(3/2)} + 2
T_{\overline{K}\Sigma}^{(1/2)}],&\nonumber\\
&T_{\overline{K}N}^{(1)}=\frac13[T_{\pi \Xi}^{(3/2)}+ 2T_{\pi
\Xi}^{(1/2)}] =\frac13[ T_{K\Sigma}^{(3/2)} + 2
T_{K\Sigma}^{(1/2)}],&\nonumber\\
&T_{\overline{K}N}^{(1)} + T_{\overline{K}N}^{(0)}= T_{K\Xi}^{(1)} +
T_{K\Xi}^{(0)} = \frac13 [T_{\pi\Sigma}^{(2)} + 3T_{\pi\Sigma}^{(1)}
+2 T_{\pi\Sigma}^{(0)} ],&\nonumber\\
&T_{\eta\Sigma}=T_{\pi\Lambda},&\nonumber\\
&\frac12[T_{\pi\Sigma}^{(2)}+T_{\pi\Sigma}^{(1)}] +T_{\pi\Lambda} =
\frac13[ 2T_{K\Sigma}^{(3/2)}+T_{K\Sigma}^{(1/2)}]+T_{K\Lambda} =
\frac13[
2T_{\overline{K}\Sigma}^{(3/2)}+T_{\overline{K}\Sigma}^{(1/2)}]+T_{\overline{K}\Lambda} &\nonumber\\
& = \frac13[2T_{\pi N}^{(3/2)}+T_{\pi N}^{(1/2)}] +T_{\eta N} =
\frac13[2T_{\pi \Xi}^{(3/2)}+T_{\pi \Xi}^{(1/2)}] +T_{\eta \Xi}.&
\end{eqnarray}

\section{Determination of Low-Energy-Constants}\label{sec5}

There are seven unknown LECs in ${\cal L}_{\phi B}^{(2)}$,
$b_{D,F,0}$, $d_{D,F,0}$, and $d_1$. The threshold T-matrices
depend on four combinations of them: $(d_F-2b_F)$, $(d_D-2b_D)$,
$(d_0-2b_0)$ and $d_1$. After the regrouping, we defined another
four combinations: $C_1$, $C_0$, $C_\pi$ and $C_d$. The values of
$C_1$, $C_0$ and $C_\pi$ can be extracted from the experimental
values of $T_{KN}$ and $T_{\pi N}$. With $C_1$, $C_0$ and $C_\pi$
one can predict most of pion-baryon and kaon-baryon scattering
lengths.

Threshold T-matrices $T_{K\Sigma}^{(1)}$, $T_{K\Sigma}^{(0)}$,
$T_{M\Lambda}$ and $T_{\eta B}$ depend on additional LECs $C_d$
and $b_{D,F}$. The values of $b_D$ and $b_F$ can be extracted from
the octet baryon mass differences. In order to determine
$C_d=d_1-\frac{D^2-3F^2}{6M_0}$, one has to fix both $d_1$ and
$M_0$.

For the sake of comparison, let's recall the procedure of
determining LECs in Ref. \cite{KN}. (1) First, Kaiser fixed the
energy scale parameter from the loop integration $\lambda$ to be
0.95 GeV using the experimental value of $T_{\pi N}^-$; (2) Then
he determined the LECs $C_{1,0}$ using the experimental values of
$T_{KN}^{(1)}$ and $T_{KN}^{(0)}$ \cite{TKN}: $C_1=2.33
\,\rm{GeV}^{-1}$, $C_0=0.36 \,\rm{GeV}^{-1}$. Throughout his
analysis he used $D=0.8$ and $F=0.5$; (3) The three LECs $b_D =
0.042 \,\rm{GeV}^{-1}$, $b_F=-0.557 \,\rm{GeV}^{-1}$, $b_0 =
-0.789 \,\rm{GeV}^{-1}$ and $M_0=918.4$ MeV were obtained by
fitting the octet baryon masses and the $\pi N$ sigma term
$\sigma_{\pi N}=45$ MeV from Ref. \cite{gasser}; (4) Using the
experimental value of $T_{\pi N}^+$, the extracted parameters
$d_F=-0.968 \,\rm{GeV}^{-1}$, $2d_0+d_D= -1.562 \,\rm{GeV}^{-1}$,
$d_1+d_D= 1.150 \,\rm{GeV}^{-1}$ and the value of $d_0\simeq -1.0
\,\rm{GeV}^{-1}$ determined elsewhere \cite{d0}, one extracts
$d_1$.

Our procedure is slightly different from Kaiser's. We take
$\lambda=4\pi f_\pi$ as the chiral symmetry breaking scale, which
is widely adopted in the chiral perturbation theory. The other
steps are somehow similar. The meson masses and decay constants
are from PDG \cite{pdg}: $m_\pi=139.57\mathrm{MeV}$,
$m_K=493.68\mathrm{MeV}$, $m_\eta =547.75\mathrm{MeV}$,
$f_\pi=92.4\mathrm{MeV}$, $f_K=113\mathrm{MeV}$, $f_\eta=1.2
f_\pi$. $D=0.8$ and $F=0.5$.

We first determine the combinations $C_1$ and $C_0$. Using the
experimental values of $a_{KN}^{(1)}=-0.33\,{\rm fm}$ and
$a_{KN}^{(0)}=0.02 \,{\rm fm}$ \cite{TKN}, we extract
\begin{eqnarray}
C_1=1.786 \,\mathrm{GeV}^{-1}, \qquad C_0=0.413 \,\mathrm{GeV}^{-1}.
\end{eqnarray}

$a_{\pi N}^+$ from one recent experiment is $-0.0001\pm
_{0.0021}^{0.0009} m_\pi^{-1}$ \cite{TpiN}, from which
$C_\pi=0.096^{+0.020}_{- 0.048} \,\mathrm{GeV}^{-1}$ is extracted.
The new value of $a_{\pi N}^+$ indicates a larger $\sigma_{\pi N}$
\cite{TpiN}. Theoretical calculations of $\sigma_{\pi N}$ lie in a
large range \cite{sigma}. For comparison, we use two values
$\sigma_{\pi N}=45 \,\mathrm{MeV}$ and 57 MeV for the following
calculation. The latter one is compatible with Ref. \cite{TpiN}. We
use $\sigma_{\pi N}=45 \,\mathrm{MeV}$ to illustrate the fitting
procedure.

We now determine $b_D$, $b_F$, $b_0$ and $M_0$ with the formulas
of the octet baryon masses and $\sigma_{\pi N}$ given in Ref.
\cite{review}. We use $f=f_\pi$ in the $\pi$ loops, $f=f_K$ in
kaon loops and $f=f_\eta$ in $\eta$ loops in these formulas. By
fitting these four parameters to baryon masses $M_N=938.9\pm 1.3$
MeV, $M_\Sigma=1193.4\pm 8.1$ MeV, $M_\Xi=1318.1 \pm 6.7$ MeV,
$M_\Lambda=1115.7\pm 5.4$ MeV and $\sigma_{\pi N}=45 \pm 8
\,\mathrm{MeV}$ \cite{gasser}, we get
\begin{eqnarray}
M_0=820.00\pm104.24 \,\mathrm{MeV}, \quad b_0=-0.819\pm 0.103
\,\mathrm{GeV}^{-1},\nonumber\\
b_D=0.044\pm 0.008 \,\mathrm{GeV}^{-1}, \quad b_F=-0.491\pm 0.003
\,\mathrm{GeV}^{-1}
\end{eqnarray}
with $\chi^2/\mathrm{d.o.f.}\sim 0.98$. Both the up and down quark
mass difference and electromagnetic interaction contribute to the
baryon mass splitting within an isospin multiplet. The typical
electromagnetic correction is roughly around 0.5\% of the baryon
mass, which is not considered in this work. Therefore we have
added some uncertainty to $M_\Lambda$. We take the central value
of $M_N$, $M_\Sigma$ and $M_\Xi$ to be the average of the isospin
multiplet. The corresponding error is simply the mass splitting of
the isospin multiplet. We determine other LECs and perform
numerical evaluations with the above LECs. The values we get
differ from those in Ref. \cite{KN} because we use a different
$f_\eta$.

From the above determined $C_{1,0,\pi}$, $M_0$, $b_{0,D,F}$, we have
\begin{eqnarray}
d_F=-0.642^{+0.038}_{-0.057} \,\mathrm{GeV}^{-1}, \qquad
2d_0+d_D=-1.722^{+0.413}_{-0.416} \,\mathrm{GeV}^{-1}, \qquad
d_1+d_D=0.689^{+0.026}_{-0.050} \,\mathrm{GeV}^{-1}.
\end{eqnarray}
We derived the errors with the standard error propagation formula.
The three LECs $d_D$, $d_0$ and $d_1$ cannot be determined
independently in the procedure. Of these LECs, $d_0$ may have the
minimum uncertainty. It is convenient to use it as an input. Up to
now, only the second order value is available. With $C_1$, $C_0$,
$C_\pi$, $M_0$, $b_0$, $b_D$ and $d_0=-0.996 \,\mathrm{GeV}^{-1}$
\cite{d0}, we get
$d_D=0.270^{+0.413}_{-0.416}\,\mathrm{GeV}^{-1},d_1=0.419^{+0.414}_{-0.423}
\,\mathrm{GeV}^{-1}$ and
$C_d=0.441^{+0.414}_{-0.423}\,\mathrm{GeV}^{-1}$.

We denote the parameter set corresponding to $\sigma_{\pi N}=45 (57)
\pm 8 \,\mathrm{MeV}$ as Set I (II) respectively, which is collected
in Table \ref{sumpara}.
\begin{table}
\begin{tabular}{|c|cccccccccc|}
\hline
&$\sigma_{\pi N}$&$C_1$&$C_0$&$C_\pi$&$M_0$  &$b_0$ &$d_D$  &$d_F$ &$d_1$ &$C_d$\\
\hline
Set 1  &$45\pm 8$       &1.786&0.413&$0.096^{+0.020}_{-0.048}$&820.00$\pm$104.24 &$-0.819\pm0.103$& $0.270^{+0.413}_{-0.416}$ &$-0.642^{+0.038}_{-0.057}$& $0.419^{+0.414}_{-0.423}$&$0.441^{+0.414}_{-0.423}$\\
\hline
Set 2  &$57\pm 8$       &1.786&0.413&$0.096^{+0.020}_{-0.048}$&663.86$\pm$104.26 &$-0.973\pm0.102$&$-0.282^{+0.415}_{-0.418}$ &$-0.585^{+0.052}_{-0.067}$& $0.974^{+0.416}_{-0.425}$&$1.001^{+0.416}_{-0.425}$\\
\hline
\end{tabular}
\caption{Two sets of parameters in the numerical analysis. Units
are MeV for $\sigma_{\pi N}$ and $M_0$ and GeV$^{-1}$ for other
quantities. The variation of sigma term has no effect on
$b_D=0.044\pm 0.008$ GeV$^{-1}$ and $b_F=-0.491\pm 0.003$
GeV$^{-1}$. $d_D$, $d_1$ and $C_d$ are deduced with $d_0=-0.996$
GeV$^{-1}$ \cite{d0}. The error of $C_\pi$ is from the
experimental scattering lengths. The errors of $M_0$, $b_0$, $b_D$
and $b_F$ are given by MINUIT. The errors of $d_D$, $d_F$, $d_1$
and $C_d$ are related to those of $C_\pi$, $b_0$, $b_D$, $b_F$ and
$M_0$.} \label{sumpara}
\end{table}

\section{Results and Discussions}\label{sec6}

The threshold T-matrices with the first and second set of
parameters are collected in Tables \ref{piB45}-\ref{etaB45} and
Tables \ref{piB57}-\ref{etaB57} respectively. The corresponding
scattering lengths are listed in the last column of these tables.
The scattering lengths do not change significantly with a larger
$\sigma_{\pi N}$. The T-matrices at the second order are expressed
with $C_1$, $C_0$, $C_\pi$, $C_d$, $b_D$ and $b_F$. The errors of
$C_1$ and $C_0$ can not been extracted from known experimental
sources. Therefore, all the errors in these tables are estimated
from $C_\pi$, $C_d$, $b_D$ and $b_F$ with the error propagation
formula.

Our $a_{\pi N}^{-}$ is consistent with the experimental value
0.125$_{-0.003}^{+0.001}$ fm \cite{TpiN} while our $a_{\pi
\Sigma}^{(0)}= 0.60\pm 0.04$ fm is smaller than the value $1.10\pm
0.06$ fm \cite{I0MB}.
\begin{table}
\begin{tabular}{|cccccc|}
\hline
                      &${\cal O}(p)$&${\cal O}(p^2)$&${\cal O}(p^3)$&Total&Scattering lengths\\\hline
$T_{\pi N}^+$         & 0           & $0.58^{+0.02}_{-0.04}$          & $-0.58$         & $-0.002^{+0.018}_{-0.043}$  &$-0.00014^{+0.00127}_{-0.00297}$ (input)  \\
$T_{\pi N}^-$        & 1.61         & 0             & 0.26          & 1.87     &0.13         \\
$T_{\pi N}^{(3/2)}$  & $-1.61$        & $0.58^{+0.02}_{-0.04}$         & $-0.85$         & $-1.88^{+0.02}_{-0.04}$   &$-0.130^{+0.001}_{-0.003}$   \\
$T_{\pi N}^{(1/2)}$  & 3.23         & $0.58^{+0.02}_{-0.04}$         & $-0.06$        & $3.75^{+0.02}_{-0.04}$   &$0.260^{+0.001}_{-0.003}$   \\
\hline
$T_{\pi\Sigma}^{(2)}$& $-3.23$        &0.80           &$-1.03$          &$-3.46$     &$-0.25$       \\
$T_{\pi\Sigma}^{(1)}$&  3.23        &$0.41^{+0.37}_{-0.38}$          &$-0.02$         & $3.61^{+0.37}_{-0.38}$   &$0.26\pm{0.03}$  \\
$T_{\pi\Sigma}^{(0)}$&  6.45        &$1.40^{+0.56}_{-0.57}$          & 0.59          & $8.44^{+0.56}_{-0.57}$   &$0.60\pm{0.04}$  \\
\hline
$T_{\pi\Xi}^{(3/2)}$ & $-1.61$        & 0.49          & $-1.25$         & $-2.37$    &$-0.17$       \\
$T_{\pi\Xi}^{(1/2)}$ & 3.23         & 0.49          & $-0.46$         & 3.26     &0.23        \\
\hline
$T_{\pi\Lambda}$     & 0            & $0.52^{+0.06}_{-0.07}$          & $-1.52$         & $-1.00^{+0.06}_{-0.07}$   &$-0.071^{+0.004}_{-0.005} $ \\
\hline
\end{tabular}
\caption{Pion-baryon threshold T-matrices order by order with the
parameter Set I in unit of fm. }\label{piB45}
\end{table}

Our isovector $\bar K N$ scattering length
$a_{\overline{K}N}^{(1)}=(0.40+0.36i)\,{\rm fm}$ is roughly
consistent with the empirical value (0.37+0.60i) fm \cite{TKN},
although the imaginary part is smaller. The isoscalar $\bar K N$
channel is strongly affected by the resonance $\Lambda(1405)$ which
lies below $\bar K N$ threshold. It's known long ago that it's
impossible to get a reasonable description of the isoscalar $\bar K
N$ scattering length without taking into account $\Lambda(1405)$'s
contribution in a non-perturbative way. Especially, $\Lambda(1405)$
affects the real part of $a_{\overline{K}N}^{(0)}$ dramatically. In
contrast, the imaginary part of our $a_{K^-p}=0.95+0.29i$ fm is
compatible with the recent experimental value,
$\mathrm{Im}[a_{K^-p}]=+(0.302\pm 0.135\pm 0.036)i$ fm \cite{DEAR}.
\begin{table}
\begin{tabular}{|c|ccccc|}
\hline
                      &${\cal O}(p)$&${\cal O}(p^2)$&${\cal O}(p^3)$&Total&Scattering lengths\\\hline
$T_{KN}^{(1)}$        & $-7.63$       &  6.73         & $-5.42$         & $-6.33$          &$-0.33$ (input)  \\
$T_{KN}^{(0)}$        & 0           & 1.56          & $-1.17$         & 0.38           &0.02 (input)  \\
$T_{\bar{K}N}^{(1)}$  & 3.81        & 4.14          & $-0.32+6.95i$   & $7.64+6.95i$     &$0.40+0.36i$          \\
$T_{\bar{K}N}^{(0)}$  & 11.44       &9.31           & $7.96+4.17i$    & $28.72+4.17i$    &$1.50+0.22i$          \\
\hline
$T_{K\Sigma}^{(3/2)}$ & $-3.81$        &$4.86^{+0.15}_{-0.36}$    &$-1.00+2.78 i$   &$0.04^{+0.15}_{-0.36}+2.78 i$   &$0.0024^{+0.0086}_{-0.0202}+0.16i$ \\
$T_{K\Sigma}^{(1/2)}$ &  7.63        &$3.78^{+0.08}_{-0.18}$    &$2.99+0.69i$     &$14.40^{+0.08}_{-0.18}+0.69i$    &$(0.81\pm 0.01)+0.04i$   \\
$T_{\overline{K}\Sigma}^{(3/2)}$& $-3.81$       &4.14           &$-4.61+2.78i$  &$-4.28+2.78i$   &$-0.24+0.16i$   \\
$T_{\overline{K}\Sigma}^{(1/2)}$& 7.63        &$5.22^{+0.23}_{-0.54}$           &$10.20+0.69i$   &$23.05^{+0.23}_{-0.54}+0.69i$   &$1.30^{+0.01}_{-0.03}+0.04i$   \\
\hline
$T_{K\Xi}^{(1)}$      & 3.81        & $4.86^{+0.15}_{-0.36}$          & $4.06+6.95i$    & $12.73^{+0.15}_{-0.36}+6.95i$    &$0.74^{+0.01}_{-0.02}+0.40i$  \\
$T_{K\Xi}^{(0)}$      &11.44        & $8.59^{+0.15}_{-0.36}$          & $5.12+4.17i$    & $25.16^{+0.15}_{-0.36}+4.17i$    &$1.46^{+0.01}_{-0.02}+0.24i$  \\
$T_{\bar{K}\Xi}^{(1)}$& $-7.63$       & 6.73          & $-4.66$         & $-5.56$          &$-0.32$               \\
$T_{\bar{K}\Xi}^{(0)}$& 0           & $3.00^{+0.31}_{-0.72}$          & 6.81          & $9.80^{+0.31}_{-0.72}$          &$0.57^{+0.02}_{-0.04}$   \\
\hline
$T_{K\Lambda}(T_{\overline{K}\Lambda})$& 0   & $4.88^{+1.04}_{-1.06}$          & $-1.76+6.25i$   & $3.11^{+1.04}_{-1.06}+6.25i$  &$(0.17\pm{0.06})+0.34i$  \\
\hline
\end{tabular}
\caption{Kaon-baryon threshold T-matrices order by order with the
parameter Set I in unit of fm. }\label{kaonB45}
\end{table}

The real part of our $a_{\eta N}=[(0.18\pm 0.07)+0.42i]\,
\mathrm{fm}$ is compatible with those in \cite{S0MB,length}. Our
imaginary part satisfies the requirement $\mathrm{Im}[a_{\eta
N}]\geq (0.28\pm 0.04)\,\mathrm{fm}$ derived in Ref. \cite{wilkin}
and is larger than those in Refs. \cite{etaNcompi,etahy}. Again, one
should be cautious about our number. Since HB$\chi$PT is a
perturbative approach, we have completely ignored the contribution
from the nearby $N^*(1535)$ resonance.

Our $a_{\eta\Sigma}= [(0.42\pm0.04)+0.30i]\,\rm{fm}$ is consistent
with the range for $a_{\eta\Sigma}=[(0.10\sim 1.10) +(0.35\sim
2.20)i]\,\rm{fm}$ in Ref. \cite{etahy}. Our
$a_{\eta\Lambda}=[(0.69\pm 0.11) +0.89i]\,\rm{fm}$ is also
consistent with the value $[(0.64\pm0.29) +(0.80\pm
0.30)i]\,\rm{fm}$ in Ref. \cite{etahy}. The above two imaginary
parts are both larger than that of $a_{\eta\Lambda}=[(0.50\pm 0.05)+
(0.27 \pm 0.01)i]\,\rm{fm}$ in Ref. \cite{I0MB}.
\begin{table}
\begin{tabular}{|c|cccc|}
\hline
                  &${\cal O}(p^2)$&${\cal O}(p^3)$&Total     &Scattering lengths\\\hline
$T_{\eta N}$          & $1.22^{+1.33}_{-1.37}$      & $2.40+8.32i$   & $3.63^{+1.33}_{-1.37}+8.32i$  &$(0.18\pm 0.07)+0.42i$\\
\hline
$T_{\eta\Sigma}$      & $5.78^{+0.68}_{-0.75}$      & $1.93+5.55i$   & $7.70^{+0.68}_{-0.75}+5.55i$   &$(0.42\pm0.04)+0.30i$ \\
\hline
$T_{\eta\Xi}$         & $10.98^{+1.34}_{-1.39}$     & $0.77+8.32i$   &$11.75^{+1.34}_{-1.39}+8.32i$  &$(0.66\pm0.08)+0.47i$\\
\hline
$T_{\eta\Lambda}$     & $10.46^{+2.00}_{-2.04}$     & $2.39+16.64i$  & $12.85^{+2.00}_{-2.04}+16.64i$ &$(0.69\pm 0.11)+0.89i$\\
\hline
\end{tabular}
\caption{Eta-baryon threshold T-matrices order by order with the
parameter Set I in unit of fm. }\label{etaB45}
\end{table}

In the SU(3) HB$\chi$PT approach, the convergence of the chiral
expansion is a serious issue because of the large mass of kaon and
eta mesons. One has to investigate case by case to make sure whether
the higher order chiral corrections converge or blow up. From Tables
\ref{piB45}-\ref{etaB45} we find the chiral corrections to the
threshold T-matrices converge well only in the following few
channels: $T_{\pi N}^{(1/2)}$, $T_{\pi\Sigma}^{(1)}$,
$T_{\pi\Sigma}^{(0)}$, $T_{\pi\Xi}^{(1/2)}$, $T_{K\Sigma}^{(1/2)}$,
$T_{K\Xi}^{(0)}$. HB$\chi$PT predictions of scattering lengths in
these channels should be reliable. $T_{\pi\Sigma}^{(1)}$ is
particular interesting. The chiral corrections converge very fast in
this channel. The LEC $C_d$ can be extracted if the scattering
length $a_{\pi\Sigma}^{(1)}$ is measured experimentally.

\begin{table}
\begin{tabular}{|c|ccccc|}
\hline
                      &${\cal O}(p)$&${\cal O}(p^2)$&${\cal O}(p^3)$&Total&Scattering lengths\\
\hline
$T_{\pi\Sigma}^{(1)}$&  3.23        &$-0.10^{+0.37}_{-0.38}$      &$-0.02$   & $3.11^{+0.37}_{-0.38}$    &$0.22\pm{0.03}$   \\
$T_{\pi\Sigma}^{(0)}$&  6.45        &$2.16^{+0.56}_{-0.57}$       & 0.59    & $9.19^{+0.56}_{-0.57}$    &$0.66 \pm 0.04$   \\
\hline
$T_{\pi\Lambda}$     & 0            & $0.60^{+0.06}_{-0.07}$      & $-1.52$   & $-0.92^{+0.06}_{-0.07}$    &$-0.065^{+0.004}_{-0.005}$  \\
\hline
\end{tabular}
\caption{Pion-baryon threshold T-matrices order by order with the
parameter Set II in unit of fm. }\label{piB57}
\end{table}

\begin{table}
\begin{tabular}{|c|ccccc|}
\hline
                      &${\cal O}(p)$&${\cal O}(p^2)$&${\cal O}(p^3)$&Total       &Scattering length \\
\hline
$T_{K\Lambda}(T_{\overline{K}\Lambda})$& 0   & $3.47^{+1.04}_{-1.07}$       & $-1.76+6.25i$   & $1.71^{+1.04}_{-1.07}+6.25i$  &$(0.09\pm0.06)+0.34i$ \\
\hline
\end{tabular}
\caption{Kaon-baryon threshold T-matrices order by order with the
parameter Set II in unit of fm. }\label{kaonB57}
\end{table}

\begin{table}
\begin{tabular}{|c|cccc|}
\hline
                      &${\cal O}(p^2)$&${\cal O}(p^3)$&Total       &Scattering lengths \\\hline
$T_{\eta N}$          & $-0.57^{+1.34}_{-1.37}$         & $2.40+8.32i$   & $1.83^{+1.34}_{-1.37}+8.32i$  &$(0.09\pm 0.07)+0.42i$    \\
\hline
$T_{\eta\Sigma}$      & $6.67^{+0.68}_{-0.75}$         & $1.93+5.55i$    & $8.60^{+0.68}_{-0.75}+5.55i$   &$(0.47\pm 0.04)+0.30i$   \\
\hline
$T_{\eta\Xi}$         & $9.19^{+1.34}_{-1.40}$         & $0.77+8.32i$  & $9.95^{+1.34}_{-1.40}+8.32i$ &$(0.56 \pm 0.08)+0.47i$       \\
\hline
$T_{\eta\Lambda}$     & $13.16^{+2.00}_{-2.05}$         & $2.39+16.64i$   & $15.55^{+2.00}_{-2.05}+16.64i$ &$(0.83\pm 0.11)+0.89i$  \\
\hline
\end{tabular}
\caption{Eta-baryon threshold T-matrices order by order with the
parameter Set II in unit of fm. }\label{etaB57}
\end{table}

We calculate the scattering lengths at threshold where chiral
perturbation theory, in principle, works well. However, some
corrections may improve the calculation. First, the complete
determination of the LECs in the second and third order Lagrangians
may give more accurate predictions in the HB$\chi$PT framework.
Secondly, due to the complicate convergence in the SU(3) case,
higher order corrections may be significant. Future investigations
to the fourth order will give us a clearer picture. Thirdly,
subthreshold effects of closed channels may give corrections.

One can not consider effects from resonances close to thresholds in
the current method. To extend the range for chiral expansion and
include resonance effects, unitarized chiral perturbation theory was
developed. The unitarity corrections to scattering lengths for $\pi
N$ scattering had been studied in Refs. \cite{ucsl} with such a
method. One may also consider coupled channel effects like the
treatment in Ref. \cite{ccue} with this method. To consider the
resonance effects from t and u channels for $\pi N$ scattering, a
more complete treatment on the left-cut is possible because there
the chiral expansion has been carried out to ${\cal O}(p^4)$.

In summary, we have calculated the chiral corrections to the s-wave
meson-baryon scattering lengths to the third chiral order in SU(3)
HB$\chi$PT. Hopefully these explicit expressions of chiral
corrections will be helpful to the chiral extrapolations of the
scattering lengths in the future lattice simulations. This is the
main result of this paper. There is a good possibility of measuring
these meson-baryon scattering lengths experimentally from the
strangeness program at CSR, Lan-Zhou and JHF in the near future.
Therefore we have also done some numerical analysis of these
scattering lengths based on the available experimental information.
We find that the chiral expansion converges quite well in several
channels. Hence, HB$\chi$PT predictions of these scattering lengths
are reliable, which may be useful to the construction of the
meson-baryon interaction models.

\section*{Acknowledgments}

This project was supported by the National Natural Science
Foundation of China under Grants 10375003, 10421503 and 10625521,
Ministry of Education of China, FANEDD, Key Grant Project of Chinese
Ministry of Education (NO 305001) and SRF for ROCS, SEM. Y.R.L.
thanks N. Kaiser for very helpful discussions. Y.R.L. thanks Y. Cui
for checking part of the calculation.

\end{document}